\documentclass[usenatbib]{mnras}
\usepackage{newtxtext,newtxmath}

\usepackage[T1]{fontenc}

\DeclareRobustCommand{\VAN}[3]{#2}
\let\VANthebibliography\thebibliography
\def\thebibliography{\DeclareRobustCommand{\VAN}[3]{##3}\VANthebibliography}

\usepackage{graphicx}
\usepackage{amsmath}

\usepackage{amssymb}
\usepackage{ulem}
\usepackage{lineno}
\usepackage{footnote}
\usepackage{color}
\title[A Fast Venusian Cloud Model]{A Fast, Semi-analytical Model for the Venusian Binary Cloud System}

\author[Dai et al.]{
L.-K. Dai$^{1}$\thanks{E-mail: dailk@mail2.sysu.edu.cn},
X. Zhang$^{2}$,
and J. Cui$^{1,3}$
\\
$^{1}$Planetary Environmental and Astrobiological Research Laboratory (PEARL), School of Atmospheric Sciences, Sun Yat-sen University, Zhuhai, Guangdong, \\People's Republic of China\\
$^{2}$Department of Earth and Planetary Sciences, University of California Santa Cruz, Santa Cruz, CA, USA\\
$^{3}$Center for Excellence in Comparative Planetology, Chinese Academy of Sciences, Hefei, Anhui, People's Republic of China
}

\pubyear{2022}

\begin{document}
\label{firstpage}
\pagerange{\pageref{firstpage}--\pageref{lastpage}}
\maketitle

\begin{abstract}

The Venusian clouds originate from the binary condensation of H$_{2}$SO$_{4}$ and H$_{2}$O. The two components strongly interact with each other via chemistry and cloud formation. Previous works adopted sophisticated microphysical approaches to understand the clouds. Here we show that the observed vapor and cloud distributions on Venus can be well explained by a semi-analytical model. Our model assumes local thermodynamical equilibrium for water vapor but not for sulfuric acid vapor, and includes the feedback of cloud condensation and acidity to vapor distributions. The model predicts strong supersaturation of the H$_{2}$SO$_{4}$ vapor above 60 km, consistent with our recent cloud condensation model. The semi-analytical model is 100 times faster than the condensation model and 1000 times faster than the microphysical models. This allows us to quickly explore a large parameter space of the sulfuric acid gas-cloud system. We found that the cloud mass loading in the upper clouds has an opposite response of that in the lower clouds to the vapor mixing ratios in the lower atmosphere. The transport of water vapor influences the cloud acidity in all cloud layers while the transport of sulfuric acid vapor only dominates in the lower clouds. This cloud model is fast enough to be coupled with the climate models and chemistry models to understand the cloudy atmospheres of Venus and Venus-like extra-solar planets.

\end{abstract}


\section{Introduction}
\label{sec1}

The globally distributed clouds on Venus are mainly composed of concentrated sulfuric acid \citep[H$_{2}$SO$_{4}$,][]{esposito1983clouds,1985P&SS...33..109K}. Their formation is governed by complex microphysical processes \citep[][]{2014Icar..231...83G, MicrophysicsofVenusianCloudsinRisingTropicalAir, 1997Icar..129..147J, 2017EP&S...69..161M, 2007Icar..191....1M, 2015P&SS..113..205P, 1977JAtS...34..417R, 1977JAtS...34..405R}, such as condensation, evaporation, nucleation, coagulation, as well as coalescence. Moreover, the concentration of sulfuric acid and water inside the cloud droplets determines the cloud acidity, which in turn strongly impacts the saturation vapor pressure of both species. Thus, all four components (the vapor and condensed phases of both sulfur acid and water) need to be solved simultaneously, further complicating the calculation. A recent model \citep[][hereafter D22]{2022JGRE..12707060D} found that the condensation and evaporation processes are sufficient to explain the observed vertical distributions of sulfuric acid and water vapor mixing ratios, cloud mass loading of the Mode-2 particles, and cloud acidity on Venus. The success of the condensation model implies that the vapor and cloud distribution simulation in previous microphysical models can be greatly simplified. However, all the aforementioned models, including D22, adopted the time-stepping method to solve the cloud formation process to achieve the steady state. With reasonable assumptions, we present here a semi-analytical model to directly obtain the steady-state solution of the cloud system on Venus without time stepping, allowing the orders-of-magnitude increase of the computational efficiency. 

Our work is built on a seminal study in \citet[][]{1994Icar..109...58K} that was later updated by \citet[][]{2015Icar..252..327K}. We call their model KP94 hereafter. The KP94 model is a semi-analytical model without time stepping. It mainly focused on H$_{2}$SO$_{4}$ and H$_{2}$O vapors in the middle and lower cloud region (below 60 km). The aerosol mass can be derived after the vapor distributions are obtained. There are two important assumptions in KP94: hydrogen flux conservation and local thermodynamic equilibrium (LTE). The sulfuric acid vapor is formed via cloud-top chemistry by sulfuric oxides and water vapor that are diffused from below. The first assumption thus states that the net upward hydrogen flux in the water (including both vapor and droplets) should be balanced by the net downward hydrogen flux in the sulfuric acid at any given altitude. The second assumption states that both sulfuric acid and water vapor follow the saturation vapor curves that are determined by the temperature and local cloud acidity. With these two assumptions, the KP94 model can directly solve for the cloud acidity and the vapor distribution given the temperature profile. The model results are consistent with the observed vapor and acidity below 60 km.

The KP94 model can only simulate the region below the sulfuric acid production (<60 km) where the LTE assumptions are generally valid. \citet[][]{2011Icar..215..197K} tried to justify the LTE assumption up to 100 km by the comparison between timescales of condensation and atmospheric mixing, but they did not take into account the H$_{2}$SO$_{4}$ chemical production. In fact, the LTE assumption breaks down for H$_{2}$SO$_{4}$ above 60 km. Using the production rates from various chemical models \citep[][]{2012Icar..218..230K, 2012Icar..217..714Z, 2020JGRE..12506195S}, \citet[][]{2022JGRE..12707060D} simulated the cloud condensation and found that H$_{2}$SO$_{4}$ is highly supersaturated above 60 km because the condensation is too slow to compete with the H$_{2}$SO$_{4}$ vapor production. Since the H$_{2}$SO$_{4}$ vapor is not in LTE, one must consider the condensation process that critically depends on the cloud mass, the droplet size, the vapor supersaturation, etc. But the KP94 model did not simultaneously simulate the clouds and their feedback to the condensation of H$_{2}$SO$_{4}$ vapor. On the other hand, \citet[][]{2022JGRE..12707060D} confirmed that the water vapor generally follows its saturation vapor profile in the upper cloud layer. Thus, the LTE assumption remains valid for water.

In this study, we developed a semi-analytical model that extends to 80 km to include the H$_{2}$SO$_{4}$ vapor production region. We kept the LTE assumption for the water vapor as in KP94 but did not assume the LTE condition for the H$_{2}$SO$_{4}$ vapor. Instead, we explicitly included the cloud formation and its feedback to the H$_{2}$SO$_{4}$ vapor abundances. Our model is more complicated than KP94 but we can still obtain a steady-state solution without time stepping. Since both the complicated microphysics and the time integration are avoided, the semi-analytical model is significantly faster than the D22 model and previous full microphysics models \citep[][]{2014Icar..231...83G, 2017EP&S...69..161M, 2007Icar..191....1M, 2015P&SS..113..205P}. In Section~\ref{sec2} we describe the equations and numerical procedure of the model. In Section~\ref{sec3} we show the results of our nominal simulation and explore a large parameter space of the H$_{2}$SO$_{4}$-H$_{2}$O system. We conclude this study in Section~\ref{sec4}. 

\section{Model}
\label{sec2}

\subsection{Hydrogen Flux Conservation}
\label{sec2.1}

\begin{figure*}
\center
\includegraphics[width=40pc]{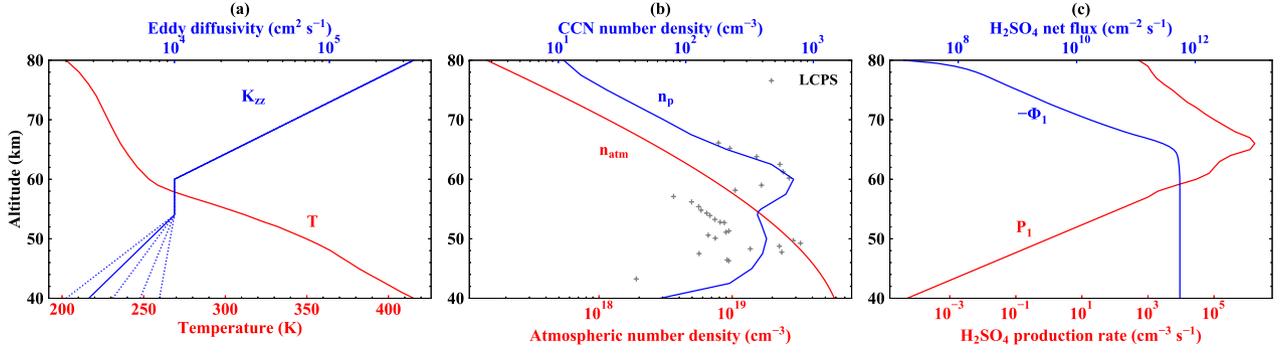}
\caption{The standard vertical profiles of the background atmospheric properties in our model: (a) the temperature profile (red) from \citet[][]{1985AdSpR...5k...3S} and the eddy diffusivity profile (blue solid line) from \citet[][]{2022JGRE..12707060D}. The eddy diffusivity profiles tested in Section~\ref{sec3.2} are represented as the blue dashed lines. We increase (or decrease) the standard diffusivity profile below 54 km by varying the diffusivity at 40 km to 2000, 4000, 6000, and 8000 cm$^{2}$ s$^{-1}$, respectively, while no variation is made above 54 km; (b) the atmospheric number density profile (red) from \citet[][]{2012Icar..218..230K} and the cloud condensation nuclei (CCN) number density profile (blue) from \citet[][]{2014Icar..231...83G}. The observations of CCN number density (grey pluses) are from the cloud particle size spectrometer (LCPS) aboard the Pioneer Venus probes \citep[][]{1980JGR....85.8039K}; (c) the H$_{2}$SO$_{4}$ chemical net production rate profile (red) from \citet[][]{2012Icar..218..230K} and the derived H$_{2}$SO$_{4}$ net flux profile (blue). Minus indicates a downward flux.}
\label{Fig1}
\end{figure*}

Following KP94, the hydrogen is transported by vapor diffusion as well as sedimentation and diffusion in both H$_{2}$SO$_{4}$ and H$_{2}$O condensates. The vertical fluxes are written as

\begin{equation}
\begin{aligned}
\Phi_{1}=-K_{zz}n_{atm}\frac{dq^{g}_{1}}{dz}+&q^{c}_{1}n_{atm}v-\\
&\frac{q^{c}_{1}}{q^{c}_{1}+q^{c}_{2}}K_{zz}n_{atm}\frac{d(q^{c}_{1}+q^{c}_{2})}{d z},
\end{aligned}
\label{eq1}
\end{equation}

\begin{equation}
\begin{aligned}
\Phi_{2}=-K_{zz}n_{atm}\frac{dq^{g}_{2}}{dz}+&q^{c}_{2}n_{atm}v-\\
&\frac{q^{c}_{2}}{q^{c}_{1}+q^{c}_{2}}K_{zz}n_{atm}\frac{d(q^{c}_{1}+q^{c}_{2})}{d z},
\end{aligned}
\label{eq2}
\end{equation}
where the subscripts $1$ and $2$ represent H$_{2}$SO$_{4}$ and H$_{2}$O, respectively, the superscripts $g$ and $c$ represent gas and condensate, respectively, $q$ represents the molar mixing ratio (with respect to the atmospheric density $n_{atm}$), $z$ represents the altitude, $\Phi_{i}$ is the vertical net flux of species $i$, $K_{zz}$ is the eddy diffusivity, $v$ is the sedimentation velocity. Since \citet[][]{2022JGRE..12707060D} suggested that the condensation of mode-2 particles was able to explain the main observed cloud properties including its acidity and abundance, we neglect the particle size distribution in our simple model here. At any altitude, the single-mode cloud particles share the same acidity and diameter, and consequently the same sedimentation velocity. The standard vertical profiles of $K_{zz}$ and $n_{atm}$ for Venus are shown in Fig.~\ref{Fig1}.  

Since the chemical reaction SO$_{3}$+2H$_{2}$O=H$_{2}$SO$_{4}$+H$_{2}$O is the main source of H$_{2}$SO$_{4}$ and the main sink of H$_{2}$O \citep[][]{2020JGRE..12506159B, 2012Icar..218..230K, 1981Natur.292..610K, 1998PhDT........17M, 2020JGRE..12506195S, 1980JGR....85.7849W, 1982Icar...51..199Y, 2015P&SS..113..205P}, the vertical transports of the hydrogen is conserved. We have:

\begin{equation}
\Phi_{1}=-\int^{\infty}_{z}P_{1}(z)dz=-\Phi_{2},
\label{eq3}
\end{equation}
where $P_{1}$ is the chemical net production rate of H$_{2}$SO$_{4}$ adopted from the chemical model in \citet[][]{2012Icar..218..230K}. Both the standard vertical profiles of $P_{1}$ and $-\Phi_{1}$ are shown in Fig.~\ref{Fig1}. 

\subsection{LTE for H$_{2}$O and Non-LTE for H$_{2}$SO$_{4}$}
\label{sec2.2}

The LTE condition states that the vapor should follow its saturation vapor mixing ratio (SVMR, $q^{svp}_{i}$) in the cloud layer. The SVMR depends on the local temperature and the cloud acidity (see Section~\ref{sec2.4}). As shown in \citet[][]{1980JGR....85.8039K}, H$_{2}$O condensation is sufficiently fast to keep its curve following the SVMR and maintain an LTE condition such that:

\begin{equation}
q^{g}_{2}=q^{svp}_{2}.
\label{eq4}
\end{equation}
But for H$_{2}$SO$_{4}$, LTE is not valid above 60 km, where the chemical production rate is high \citep[][]{1980JGR....85.8039K}. The vapor abundance is mainly controlled by the vapor condensation and the chemical production, while the vapor diffusion is negligible \citep[][]{2022JGRE..12707060D}. To account for this non-LTE effect, we introduce a vapor condensation rate that balances the net chemical production of the H$_{2}$SO$_{4}$ vapor: 

\begin{equation}
P_{1}=C_{1}(q^{g}_{1}-q^{svp}_{1}),
\label{eq5}
\end{equation}
where $C_{1}$ is the condensation rate coefficient that critically depends on the cloud particle size and their abundances (see Section~\ref{sec2.3}). The term $P_{1}/C_{1}=q^{g}_{1}-q^{svp}_{1}$ quantifies the supersaturation of the H$_{2}$SO$_{4}$ vapor.

We introduce the H$_{2}$O/H$_{2}$SO$_{4}$ molar ratio in the droplets $m=q^{c}_{2}/q^{c}_{1}$. Substituting equations~(\ref{eq3}-\ref{eq5}) to equations~(\ref{eq1}-\ref{eq2}), we can obtain:

\begin{equation}
\begin{aligned}
-\int^{\infty}_{z}P_{1}(z)dz=&-K_{zz}n_{atm}\frac{d\left(q^{svp}_{1}+\frac{P_{1}}{C_{1}}\right)}{dz}+q^{c}_{1}n_{atm}v-\\
&\frac{1}{1+m}K_{zz}n_{atm}\frac{d[(1+m)q^{c}_{1}]}{dz},
\label{eq6}
\end{aligned}
\end{equation}

\begin{equation}
\begin{aligned}
\int^{\infty}_{z}P_{1}(z)dz=&-K_{zz}n_{atm}\frac{dq^{svp}_{2}}{dz}+q^{c}_{2}n_{atm}v-\\
&\frac{m}{1+m}K_{zz}n_{atm}\frac{d\left[\left(1+\frac{1}{m}\right)q^{c}_{2}\right]}{dz}.
\label{eq7}
\end{aligned}
\end{equation}
Note that $C_{1}$, $q^{svp}_{i}$, $v$, and $m$ are functions of $q^{c}_{1}$ and $q^{c}_{2}$ (see Section~\ref{sec2.3}). Given $P_{1}$, $K_{zz}$ and $n_{atm}$, one can solve for $q^{c}_{1}$ and $q^{c}_{2}$ in (6) and (7) without time stepping.

The condensate settling and diffusional fluxes can be eliminated in (6) and (7), which further simplifies the calculation. The governing equation of $m$ can be analytically obtained:

\begin{equation}
\begin{aligned}
m=\frac{K_{zz}n_{atm}\frac{dq^{svp}_{2}}{dz}+\int^{\infty}_{z}P_{1}(z)dz}{K_{zz}n_{atm}\frac{d\left(q^{svp}_{1}+\frac{P_{1}}{C_{1}}\right)}{dz}-\int^{\infty}_{z}P_{1}(z)dz}.
\label{eq8}
\end{aligned}
\end{equation}
The main difference between our equation and equation (4) in KP94 is the supersaturation term $P_{1}/C_{1}$ that is related to the H$_{2}$SO$_{4}$ condensation. If the condensation is very fast, $C_{1}$ approaches infinity and the equation~(\ref{eq5}) is reduced to that in the KP94 model. The detailed procedure to solve the system will be given in Section~\ref{sec2.6}.

\subsection{Condensation Rate Coefficient, Cloud Particle Size and Settling Velocity}
\label{sec2.3}

The condensation rate coefficient of H$_{2}$SO$_{4}$ is adopted from \citet[][]{1980JGR....85.8039K}:

\begin{equation}
\begin{aligned}
C_{1}=\frac{2\upi n_{p}D_{1}M_{1}n_{atm}D_{p}}{(M_{1}+M_{2}m)\left(1+\frac{2\lambda_{1}}{D_{p}\alpha}\right)},
\label{eq9}
\end{aligned}
\end{equation}
where $n_{p}$ is the cloud condensation nuclei (CCN, Fig.~\ref{Fig1}) number density adopted from \citet[][]{2014Icar..231...83G}, $M_{1}=98$ g mol$^{-1}$ and $M_{2}=18$ g mol$^{-1}$ are the relative molecular weights, $D_{p}$ is the particle diameter. $\alpha=1$ is the accommodation coefficient \citep[][p. 500-502]{seinfeld2016atmospheric}, $D_{1}$ is the molecular diffusivity of H$_{2}$SO$_{4}$ \citep[][]{1977JAtS...34.1104H}:

\begin{equation}
\begin{aligned}
D_{1}=\frac{\lambda_{1}}{3}\left(\frac{8kTN_{a}}{\upi M_{1}}\right)^\frac{1}{2},
\label{eq10}
\end{aligned}
\end{equation}
where $k$ is the Boltzmann constant, $N_{a}$ is the Avogadro constant, $T$ is the atmospheric temperature (Fig.~\ref{Fig1}) adopted from the observation of Pioneer Venus at 45° latitude \citep[][]{1985AdSpR...5k...3S}, $\lambda_{1}$ is the mean free path of the H$_{2}$SO$_{4}$ vapor \citep[][]{1977JAtS...34.1104H}:

\begin{equation}
\begin{aligned}
\lambda_{1}=\left[\upi n_{atm}\left(\frac{d_{1}+d_{atm}}{2}\right)^{2} \left(\frac{M_{atm}}{M_{1}+M_{atm}}\right)^\frac{1}{2} \right]^{-1},
\label{eq11}
\end{aligned}
\end{equation}
Where $d_{1}=4.4\times 10^{-8}$ cm \citep[assumed in][]{2000JPCA..104.1715H} is the molecular diameter of H$_{2}$SO$_{4}$, $d_{atm}=3.3\times 10^{-8}$ cm is the molecular diameter of the atmospheric major component \citep[carbon dioxide, CO$_{2}$,][]{2014JPCA..118.1150M}, $M_{atm}=44$ g mol$^{-1}$ is the relative molecular weight of CO$_{2}$. 

The particle size is calculated given the condensate molar mixing ratios $q^{c}_{1}$ and $q^{c}_{2}$:

\begin{equation}
\begin{aligned}
D_{p}=\left[\frac{6(M_{1}q^{c}_{1}+M_{2}q^{c}_{2})n_{atm}}{\upi n_{p}\rho_{p}N_{a}}\right]^\frac{1}{3}.
\label{eq12}
\end{aligned}
\end{equation}

The particle sediments with the Stokes velocity:

\begin{equation}
\begin{aligned}
v=-\frac{2}{9}\frac{g\rho_{p}{D_{p}}^{2}}{4\eta}C_{c},
\label{eq13}
\end{aligned}
\end{equation}
where $g=870$ cm s$^{-2}$ is the gravitational acceleration on Venus, $\eta$ is the dynamic viscosity of CO$_{2}$ gas that varies with temperature \citep[][]{1998JPCRD..27...31F}, $C_{c}$ is the Cunningham slip correction factor following \citet[][p. 371-372]{seinfeld2016atmospheric}:

\begin{equation}
\begin{aligned}
C_{c}=1+\frac{2\lambda_{1}}{D_{p}}\left(1.257+0.4e^{\frac{-1.1D_{p}}{2\lambda_{1}}}\right).
\label{eq14}
\end{aligned}
\end{equation}

\subsection{Saturation Vapor Mixing Ratios}
\label{sec2.4}

The SVMRs ($q^{svp}_{i}$) of both H$_{2}$O and H$_{2}$SO$_{4}$ are calculated by:

\begin{equation}
\begin{aligned}
\ln P^{svp}_{i}&=\ln (q^{svp}_{i}kTn_{atm})\\
&=\ln P^{svp}_{i,p}+\frac{\umu_{i}-\umu^{0}_{i}}{R_{g}T}+\frac{4\sigma M_{i}}{R_{g}T\rho_{p}D_{p}},
\label{eq15}
\end{aligned}
\end{equation}
where $P^{svp}_{i}$ is the saturation vapor pressure (SVP, in unit of bar) of species $i$ over the sulfuric acid solution, $P^{svp}_{i,p}$ is the SVP (in unit of bar) over pure species $i$, $\umu_{i}-\umu^{0}_{i}$ is the difference of the chemical potential of species $i$ between the solution and its pure condensate. This chemical potential difference varies with temperature and acidity, and the data of it in this study is derived from \citet[][]{1991JPCRD..20.1157Z}. The last term comes from the Kelvin effect \citep[][p.419-423]{seinfeld2016atmospheric}, where $\sigma=71.11$ erg cm$^{-2}$ is the surface tension of sulfuric acid \citep[][80\% sulfuric acid, 273 K]{myhre1998density}, $\rho_{p}=1.8$ g cm$^{-3}$ is the density of the droplets. The $P^{svp}_{1,p}$ is from \citet[][]{1990JChPh..93..696K}:

\begin{equation}
\begin{aligned}
\ln P^{svp}_{1,p}=&16.259-\frac{10156}{T_{0}}+\\
&10156\left[-\frac{1}{T}+\frac{1}{T_{0}}+\frac{0.38}{T_{c}-T_{0}}\left(1+\ln \frac{T_{0}}{T}-\frac{T_{0}}{T}  \right) \right],
\label{eq16}
\end{aligned}
\end{equation}
where $T_{0}=360$ K is a reference temperature, $T_{c}=905$ K is the critical temperature. The $P^{svp}_{2,p}$ is from \citet[][]{1997GeoRL..24.1931T}:

\begin{equation}
\begin{aligned}
\ln P^{svp}_{2,p}=e^{18.4524-\frac{3505.2}{T}-\frac{330918.6}{T^{2}}+
\frac{12725068.3}{T^{3}}}.
\label{eq17}
\end{aligned}
\end{equation}

\subsection{Cloud base Determination}
\label{sec2.5}

\begin{figure}
\center
\includegraphics[width=20pc]{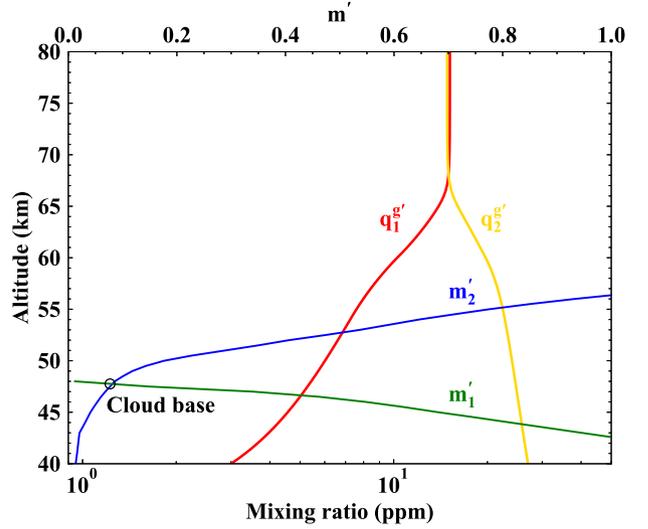}
\caption{The definition of the cloud base. The H$_{2}$SO$_{4}$ ($q^{g’}_{1}$, red) and H$_{2}$O ($q^{g’}_{2}$, yellow) vapor profiles are only controlled by their chemical production/loss and diffusion. The $m’_{1}$ profile (green) is a solution of equation~(\ref{eq18}) and the $m’_{1}$ profile (blue) is a solution of equation~(\ref{eq19}). Their intersection (black circle) represents the cloud base altitude.}
\label{Fig2}
\end{figure}

Unlike the microphysical models, our simple model (and the KP94 model) requires a known cloud base. We follow the definition of the cloud base altitude in KP94 but use a different method to find it. The cloud base altitude is where both the H$_{2}$SO$_{4}$ and H$_{2}$O vapors start to condense together. Below the cloud base, there is no condensate, and the H$_{2}$SO$_{4}$ and H$_{2}$O vapors are only controlled by the eddy diffusion and their chemical production/loss in the upper clouds. Given their mixing ratios at the lower boundary well below the cloud base (e.g., 40 km), the chemical-diffusion equilibrium vapor profiles are shown in Fig.~\ref{Fig2} ($q^{g’}_{1}$ for H$_{2}$SO$_{4}$ and $q^{g’}_{2}$ for H$_{2}$O). At the cloud base, both vapor mixing ratios should be equal to their SVMRs that are controlled by the cloud acidity (or $m$). Mathematically, one needs to find an altitude where a single value $m$ exists to satisfy both following conditions:

\begin{equation}
\begin{aligned}
q^{g'}_{1}(z)=q^{svp}_{1}(m,z),
\label{eq18}
\end{aligned}
\end{equation}

\begin{equation}
\begin{aligned}
q^{g'}_{2}(z)=q^{svp}_{2}(m,z).
\label{eq19}
\end{aligned}
\end{equation}
Note that $q^{svp}_{i}$ also depends on $z$ because the temperature changes with altitude. Graphically, one can solve for $m$ in equations~(\ref{eq18}) and (\ref{eq19}) respectively. The vertical profiles of the solution are shown in Fig.~\ref{Fig2} ($m’_{1}$ for H$_{2}$SO$_{4}$ and $m’_{2}$ for H$_{2}$O). The intersection between these two H$_{2}$O/H$_{2}$SO$_{4}$ molar ratio profiles denotes the altitude of the cloud base. One can show that the cloud base altitude obtained in our method is fully consistent with that in KP94 but our method might be more intuitively understandable.

\begin{table}
 \caption{Model parameters and boundary conditions}
 \label{table1}
 \begin{tabular}{lcc}
  \hline
  \textbf{Parameter} & \textbf{Symbol} & \textbf{Nominal model}\\
  \hline
  Atmospheric temperature & $T$ & Red line in Fig.~\ref{Fig1}(a) \\
  Eddy diffusivity & $K_{zz}$ & Solid blue line in Fig.~\ref{Fig1}(a)  \\
  Atmospheric density & $n_{atm}$ & Red line in Fig.~\ref{Fig1}(b) \\
  CCN number density & $n_{p}$  & Blue line in Fig.~\ref{Fig1}(b) \\
  H$_{2}$SO$_{4}$ chemical net production rate & $P_{1}$ & Red line in Fig.~\ref{Fig1}(c) \\
  H$_{2}$SO$_{4}$ net flux & $\Phi_{1}$ & Blue line in Fig.~\ref{Fig1}(c) \\
  Molecular diffusivity of H$_{2}$SO$_{4}$ & $D_{1}$ & equation~(\ref{eq10})  \\
  Mean free path of H$_{2}$SO$_{4}$ & $\lambda_{1}$ & equation~(\ref{eq11}) \\
  Dynamic viscosity of CO$_{2}$ & $\eta$ & \citet[][]{1998JPCRD..27...31F} \\
  Relative molecular weight of H$_{2}$SO$_{4}$ & $M_{1}$ & 98 g mol$^{-1}$ \\
  Relative molecular weight of H$_{2}$O & $M_{2}$ & 18 g mol$^{-1}$ \\
  Relative molecular weight of CO$_{2}$ & $M_{atm}$ & 44 g mol$^{-1}$  \\
  Droplet density & $\rho_{p}$ & 1.8 g cm$^{-3}$ \\
  Gravitational acceleration & $g$ & 870 cm s$^{-2}$\\
  Accommodation coefficient& $\alpha$ & 1 \\
  Surface tension of sulfuric acid & $\sigma$ & 71.11 erg cm$^{-2}$ \\
  Chemical potential & $\mu$ & \citet[][]{1991JPCRD..20.1157Z} \\
  Molecular diameter of H$_{2}$SO$_{4}$ & $d_{1}$ & 4.4$\times$10$^{-8}$ cm\\
  Molecular diameter of CO$_{2}$ & $d_{atm}$ & 3.3$\times$10$^{-8}$ cm\\
  & & \\
  \textbf{Upper boundary (80 km)}&  &\\
  H$_{2}$SO$_{4}$ vapor & $dq^{g}_{1}/dz$ & 0\\
  H$_{2}$O vapor& $dq^{g}_{2}/dz$ & 0 \\
  H$_{2}$SO$_{4}$ condensate & $dq^{c}_{1}/dz$ & 0 \\
  H$_{2}$O condensate & $dq^{c}_{2}/dz$ & 0\\
  & & \\
  \textbf{Lower boundary (40 km)}&  & \\
  H$_{2}$SO$_{4}$ vapor & $q^{g}_{1}$ & 3 ppm \\
  H$_{2}$O vapor& $q^{g}_{2}$ & 27 ppm \\
  H$_{2}$SO$_{4}$ condensate & $q^{c}_{1}$ & 0 ppm \\
  H$_{2}$O condensate & $q^{c}_{2}$ & 0 ppm \\
  \hline
 \end{tabular}
\end{table}

\subsection{Boundary condition and iterative procedure for solution}
\label{sec2.6}

This model extends from 40 to 80 km. Both vapors and condensates are assumed to be zero fluxes at the upper boundary. The H$_{2}$SO$_{4}$ and H$_{2}$O vapor mixing ratios at the lower boundary of the model domain (40 km) are fixed at 3 and 27 ppm, respectively, agreeing with the observations of Venus Express \citep[][]{2008JGRE..113.0B07M, 2012Icar..221..940O, 2021Icar..36214405O} and ground-based spectroscopy \citep[][]{1991Sci...251..547D, 1995AdSpR..15d..79D, 1993Icar..103....1P, 2014JGRE..119.1860A}. We conclude the parameters and boundary conditions in Table~\ref{table1}.

The actual lower boundary for the condensates is the cloud base defined in Section~\ref{sec2.5}. Different from the D22 model where the droplets precipitate below the cloud base and evaporate at lower altitudes, this model assumes no condensate below the cloud base. In other words, the droplets evaporate so rapidly within an infinitesimal distance that we neglect their downward diffusion at the cloud base. This assumption is fine because the droplet mass flux at the cloud base is primarily dominated by the gravitational settling of large particles. However, it does cause a small difference between our model results and that in D22, as shown in Section~\ref{sec3}.

Our procedure to solve the system follows iterative steps below:

\begin{enumerate}

\item Find out the cloud base altitude using the method in Section~\ref{sec2.5};
\item Given $C_{1}$ (for the initial $C_{1}$, one can use infinity), solve for $m$ in equation~(\ref{eq8}) from the cloud base using the central difference scheme and the scant method;
\item For each altitude, calculate $q^{g}_{1}$ and $q^{g}_{2}$ (equations~\ref{eq4}-\ref{eq5}) based on $m$ and $C_{1}$ in step (\romannumeral2);
\item Given $D_{p}$ (for the initial $D_{p}$, one can assume 1 \micron), derive sedimentation velocity $v$ using equation~(\ref{eq13});
\item Given the derived $m$ in step (\romannumeral2), $q^{g}_{i}$ in step (\romannumeral3), and $v$ in step (\romannumeral4), and assuming no particle diffusion at the cloud base, we can integrate equations~(\ref{eq6}) and (\ref{eq7}) from the cloud base to find out the profiles of $q^{c}_{1}$ and $q^{c}_{2}$ respectively;
\item Given $q^{c}_{1}$ and $q^{c}_{2}$, update $D_{p}$ using equation~(\ref{eq12});
\item Iterate steps (\romannumeral4-\romannumeral6) to get a converged $D_{p}$ profile;
\item Given $D_{p}$ in step (\romannumeral7), update $C_{1}$ using equation~(\ref{eq9});
\item Iterate steps (\romannumeral2-\romannumeral8) to find a converged solution of $m$, $q^{g}_{i}$ and $q^{c}_{i}$.
\end{enumerate}

The model reaches the steady state when the maximum variation of $m$ profile is less than 0.001\% at all altitudes. Using Python, this model is able to reach the steady state within 10 seconds. The vertical resolution is 0.5 km for the runs in this study, the same as D22 model. But the model behavior remains the same in a higher resolution.

\section{Results}
\label{sec3}

\begin{figure*}
\center
\includegraphics[width=40pc]{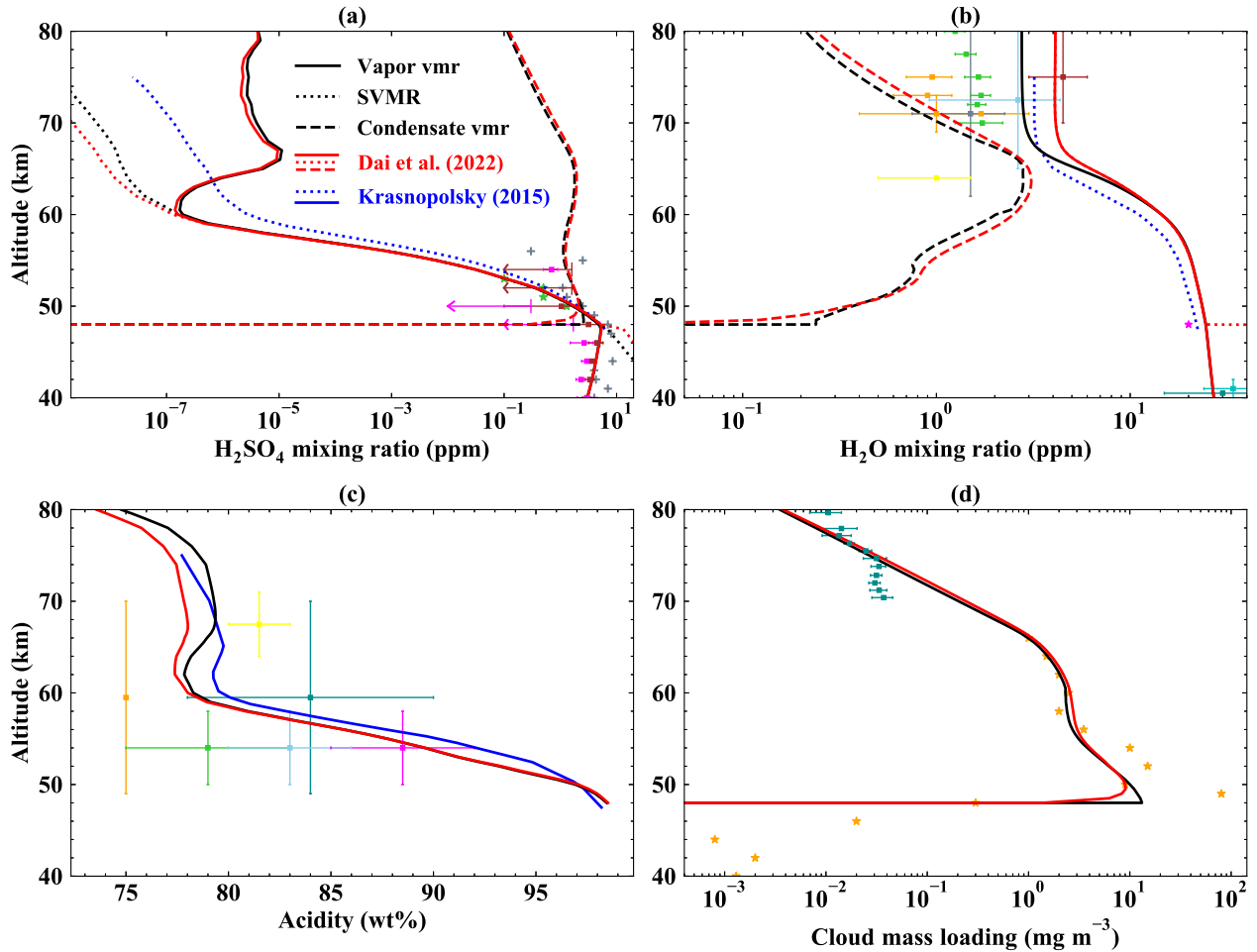}
\caption{The results of our nominal simulation and the comparisons to observations and KP94 and D22 models. (a) the volume mixing ratios of H$_{2}$SO$_{4}$ vapor (black solid line), condensate (black dashed line), and SVMR (black dotted line). The results from D22 and KP94 models are represented by red and blue lines, respectively, with the same line styles. The H$_{2}$SO$_{4}$ vapor profile above 60 km in KP94 is hardly retrieved from their linear-scale plots. Because their vapor is assumed in LTE, here we calculate the profile using their acidity and the thermodynamical data from \citet[][]{giauque1960thermodynamic}. The observations of the H$_{2}$SO$_{4}$ vapor mixing ratio are from Magellan Mission \citep[brown,][]{1998Icar..132..151K}, Venus Express (light green, \citealt{2012Icar..221..940O}; pink, \citealt{2021Icar..36214405O}), and Akatsuki Mission (gray, \citealt{2017EP&S...69..137I}); (b) the volume mixing ratios of H$_{2}$O vapor, condensate, SVMR, as well as comparisons to D22 and KP94 models. The observations are from Venus Express (orange, \citealt{2007Natur.450..646B}; light green, \citealt{2008JGRE..113.0B22F}) and ground-based spectroscopy (cyan, \citealt{1991Sci...251..547D}; dark cyan, \citealt{1995AdSpR..15d..79D}; gray, \citealt{2013A&A...559A..65E}; yellow, \citealt{2020A&A...639A..69E}; brown, \citealt{2007Icar..188..288G}; pink, \citealt{1996JGR...101.4595M}; sky blue, \citealt{2005Icar..177..129S}; (c) the distributions of the cloud acidity in our nominal simulation, D22 and KP94. The estimated values are from Venus Express (light green, \citealt{2014JGRE..119.1860A}; pink, \citealt{2012Icar..217..542B}; yellow, \citealt{2012Icar..217..561C}; sky blue, \citealt{2021PSJ.....2..153M}) and ground-based spectroscopy (orange, \citealt{1974JAtS...31.1137H}; dark cyan, \citealt{1978Icar...34...28P}); (d) the distributions of cloud mass loading from our nominal simulation and D22. The observations are from Pioneer Venus (orange, \citealt{1980JGR....85.8039K}) and Venus Express (dark cyan, \citealt{2009JGRE..114.0B42W}).}
\label{Fig3}
\end{figure*}

\subsection{Nominal model results}
\label{sec3.1}

We first compare our nominal model results with the observations and with the simulations results in other models. For the model comparison, we mainly focus on the KP94 model and the simple condensation model (D22). Our model adopts the same condensation/evaporation physics in D22 and thus the two models are very similar. The only difference, other than the numerical method, is that this semi-analytical model pre-defined a cloud base property (i.e., altitude, vapor mixing ratios, and the cloud acidity) using the method in Section~\ref{sec2.5}, while the cloud base is self-consistently simulated in the D22 condensation model. In other words, our model imposed a discontinuity of the droplet flux at the cloud base and neglected the cloud-base droplet diffusion. Because the hydrogen flux should be conserved, the cloud mass loading and the sedimentation flux at the cloud base in our model are a bit larger than those in D22 (Fig. 3). The same argument also applies to the KP94 model. Moreover, the KP94 model also assumes the LTE condition for the H$_{2}$SO$_{4}$ while our model did not. We do not compare our results to microphysical models \citep[e.g.,][]{2014Icar..231...83G, 2017EP&S...69..161M, 1977JAtS...34..417R} because the differences between the condensation-only models and the full microphysical model have been discussed in \citet[][]{2022JGRE..12707060D} in detail. 

The calculated H$_{2}$SO$_{4}$ vapor mixing ratio (Fig.~\ref{Fig3}) decreases from 5.5 ppm at the cloud base to 1.7$\times$10$^{-7}$ ppm at 60 km, then it increases to 1$\times$10$^{-5}$ ppm at 66 km. The model results agree with the observations of Magellan Mission \citep[][]{1998Icar..132..151K}, Akatsuki Mission \citep[][]{2017EP&S...69..137I}, Venus Express \citep[][]{2012Icar..221..940O, 2021Icar..36214405O}. Our model results agree pretty well with a previous simulation in the condensation model D22 using time stepping. Below 60 km, our model is also consistent with that in the KP94 model, implying that LTE is a good assumption below 60 km. Note that the KP94 used slightly different vapor mixing ratios at their lower boundary (47 km) from our model.

\begin{figure}
\center
\includegraphics[width=20pc]{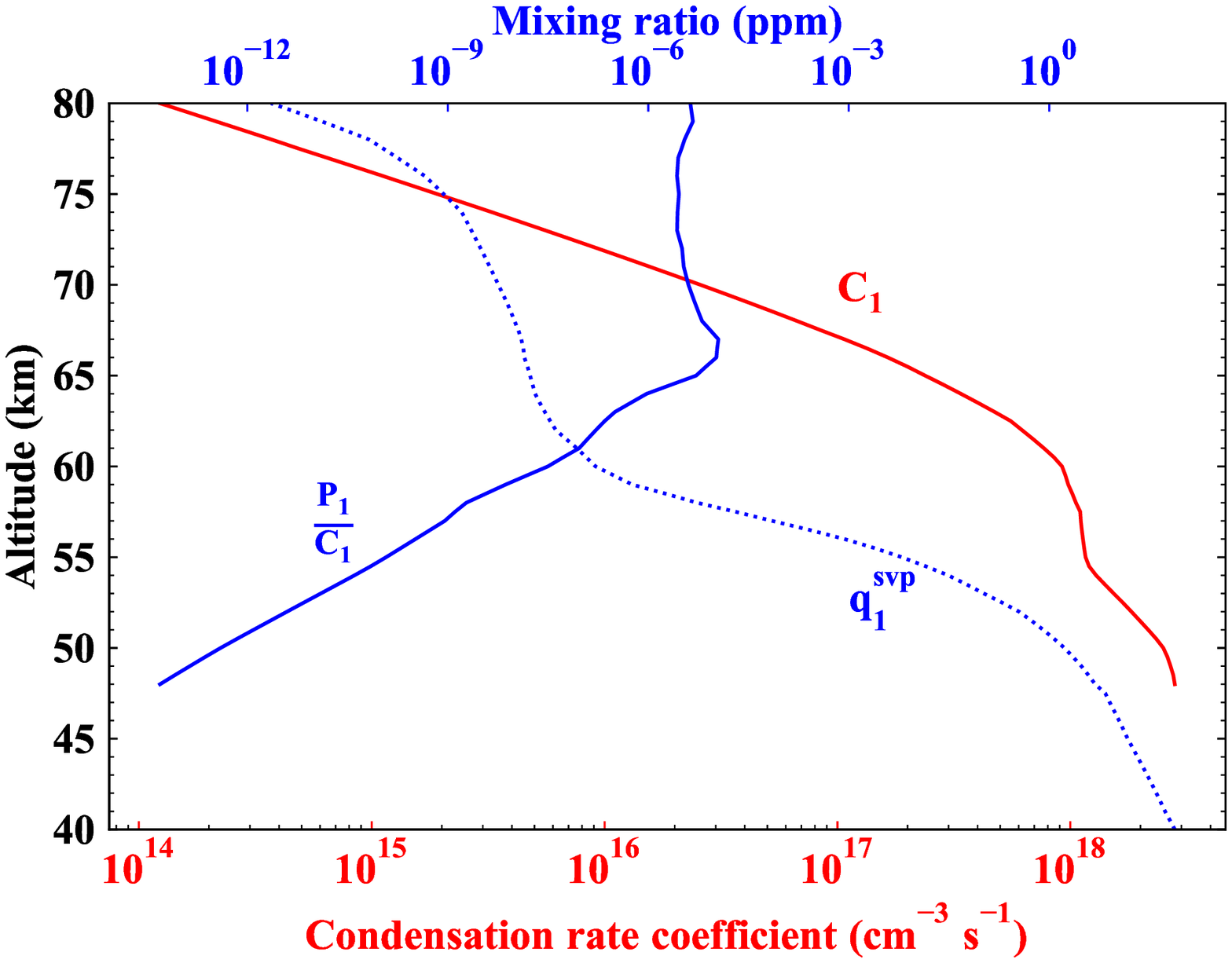}
\caption{The formation of the H$_{2}$SO$_{4}$ supersaturation, including the distributions of the H$_{2}$SO$_{4}$ condensation rate coefficient $C_{1}$ (red), the supersaturated H$_{2}$SO$_{4}$ mixing ratio $P_{1}/C_{1}$ (blue solid line), and the SVMR of H$_{2}$SO$_{4}$ $q^{svp}_{1}$ (blue dotted line).}
\label{Fig4}
\end{figure}

The H$_{2}$SO$_{4}$ vapor does not follow LTE above 60 km. Our simulated H$_{2}$SO$_{4}$ supersaturation above 60 km is consistent with that in \citet[][]{2022JGRE..12707060D}. Fig.~\ref{Fig4} illustrates the resultant H$_{2}$SO$_{4}$ condensation rate coefficient $C_{1}$, the supersaturated H$_{2}$SO$_{4}$ mixing ratio $P_{1}/C_{1}$ ($=q^{g}_{1}-q^{svp}_{1}$), and the SVMR of H$_{2}$SO$_{4}$ $q^{svp}_{1}$. The condensation rate coefficient decreases exponentially with increasing altitude. This again confirms the conclusion in \citet[][]{2022JGRE..12707060D} that a slow condensation results in a large vapor supersaturation in the upper cloud region. The supersaturated H$_{2}$SO$_{4}$ mixing ratio exceeds SVMR at 60 km, where the H$_{2}$SO$_{4}$ vapor pressure is consequently about twice the saturation vapor pressure. Above 60 km, the large chemical production of H$_{2}$SO$_{4}$ leads to an inversion of H$_{2}$SO$_{4}$ vapor profile. 

\citet[][]{2011Icar..215..197K} predicted that the supersaturation of H$_{2}$SO$_{4}$ could occur above 100 km by comparing the timescales of condensation and eddy mixing. But they did not consider the H$_{2}$SO$_{4}$ chemical production which should be much shorter. Here we give a simple estimate. The H$_{2}$SO$_{4}$ chemical net production rate at 70 km is $P_{1}$(70 km)=1$\times$10$^{5}$ cm$^{-3}$ s$^{-1}$ \citep[][]{2012Icar..218..230K}. The H$_{2}$SO$_{4}$ vapor abundance at 70 km is $n_{1}$(70 km)=1$\times$10$^{7}$ cm$^{-3}$, adopted from \citet[][]{2011Icar..215..197K}.  \citet[][]{2022JGRE..12707060D} proposed that chemical production of H$_{2}$SO$_{4}$ vapor is much stronger than its diffusional divergence at 70 km. Then its chemical timescale is $n_{1}$(70 km)/$P_{1}$(70 km)=100 s at 70 km, which is much smaller than the condensation timescale of 800 s in \citet[][]{2011Icar..215..197K}. It indicates that the condensation is not efficient enough to consume the quickly produced H$_{2}$SO$_{4}$ vapor, which leads to the supersaturation. In the upper atmosphere, the lack of CCN could also lead to a low condensation rate and subsequently induce the supersaturation \citep[][]{2017EP&S...69..161M}.

The derived H$_{2}$O vapor mixing ratio (Fig.~\ref{Fig3}) decreases from 25 ppm at the cloud base to 2.5 ppm at 70 km, agreeing with the observation of ground-based spectroscopy \citep[][]{1991Sci...251..547D, 1995AdSpR..15d..79D, 2007Icar..188..288G, 1996JGR...101.4595M, 2005Icar..177..129S}. It is about 1 ppm larger than observations of Venus Express \citep[][]{2007Natur.450..646B, 2008JGRE..113.0B22F} and ground-based spectroscopy \citep[][]{2013A&A...559A..65E, 2020A&A...639A..69E} above 70 km, but the model results could be sensitive to the lower boundary condition \citep[][]{2022JGRE..12707060D}. Our H$_{2}$O vapor mixing ratio is generally consistent with that in D22 model within 30\% (~3 ppm versus 4 ppm). As mentioned earlier, our model neglects part of the upward diffusion of the re-evaporated vapor below the cloud base. The differences with Krasnopolsky (2015) are caused by different thermodynamical properties in the SVMR calculations and different lower boundary conditions. They adopted thermodynamical data from \citet[][]{giauque1960thermodynamic}, where the temperature variations on chemical potential were not included in their study.

The vertical profile of the cloud mass loading (Fig.~\ref{Fig3}) agrees with observations of LCPS \citep[][]{1980JGR....85.8039K} and estimates based on Venus Express \citep[][]{2009JGRE..114.0B42W} except that it is underestimated by a factor of two in 52-55 km and overestimated by a factor of two in 70-74 km. The simulated mass loading is generally consistent with that in D22 model, except at the cloud base. Our calculated cloud base altitude is 48 km, with a sharp discontinuity in cloud mass, while that in the D22 model is smooth when the cloud-base particle diffusion is included. Because of this different treatment of the cloud base, the condensed H$_{2}$SO$_{4}$ is larger than their results in the lower clouds. But note that both our model and the D22 model underestimated the cloud mass by a factor of ten near the cloud base. This is because we did not include the large Mode-3 particles in our study. Based on the observations from Pioneer Venus, \citet[][]{TOON1984143} suspected that there could be large, solid particles in the lower clouds that contribute to the mass loading. \citet[][]{2021PSJ.....2..133R} proposed that the Mode-3 particles are made of hydroxide salts such as NaOH and Ca(OH)$_{2}$. Our study further confirms that H$_{2}$SO$_{4}$-H$_{2}$O system appears not sufficient to explain mass loading in the lower clouds.

The cloud acidity (Fig.~\ref{Fig3}) decreases from 98\% at the cloud base to 78\% at 62 km, then increases to 80\% at 67 km. It is consistent with the estimate of \citet[][]{2012Icar..217..542B, 2021PSJ.....2..153M, 1978Icar...34...28P} but is lower than \citet[][]{2012Icar..217..561C} in the upper clouds and higher than \citet[][]{1974JAtS...31.1137H} in the cloud deck. As mentioned above, our cloud acidity is larger above 60 km because of less water vapor flux from below. The high cloud-base acidity agrees with both the KP94 and D22 models. Our acidity in 60-66 km is smaller than KP94 model due to the different thermodynamical treatments. 

\begin{figure}
\center
\includegraphics[width=20pc]{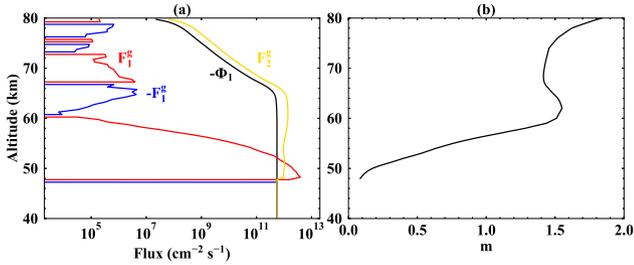}
\caption{(a) the H$_{2}$SO$_{4}$ net flux $-\Phi_{1}$ (black, minus for log scale), the H$_{2}$SO$_{4}$ vapor diffusion flux (red for $F^{g}_{1}$ and blue for $-F^{g}_{1}$), and the H$_{2}$O vapor diffusion flux ($F^{g}_{2}$, yellow); (b) the derived droplet H$_{2}$O/H$_{2}$SO$_{4}$ molar ratio $m$.}
\label{Fig5}
\end{figure}

The cloud acidity tightly interacts with vapor vertical fluxes and the H$_{2}$SO$_{4}$ net flux (see equation~\ref{eq8}). Fig.~\ref{Fig5} shows the distributions of vapor diffusion fluxes ($F^{g}_{1}$ for H$_{2}$SO$_{4}$ and $F^{g}_{2}$ for H$_{2}$O), the H$_{2}$SO$_{4}$ net flux ($\Phi_{1}$), as well as the H$_{2}$O/H$_{2}$SO$_{4}$ molar ratio in the droplets ($m$). The H$_{2}$O diffusion flux is relatively constant between 48-66 km, and potentially decreases with altitude above 66 km. Its value remains on the same order of magnitude as the H$_{2}$SO$_{4}$ net flux, indicating that eddy diffusion of H$_{2}$O vapor influences the cloud acidity over the cloud deck and the upper hazes. The H$_{2}$SO$_{4}$ vapor diffusion flux dominates in the lower clouds but rapidly decreases with altitude in the middle and upper clouds. Its value is over three orders of magnitude smaller than H$_{2}$SO$_{4}$ net flux above 60 km. This indicates that H$_{2}$SO$_{4}$ vapor diffusion largely regulates the cloud acidity in the lower clouds but has much less effect above 60 km. 

\begin{figure*}
\center
\includegraphics[width=40pc]{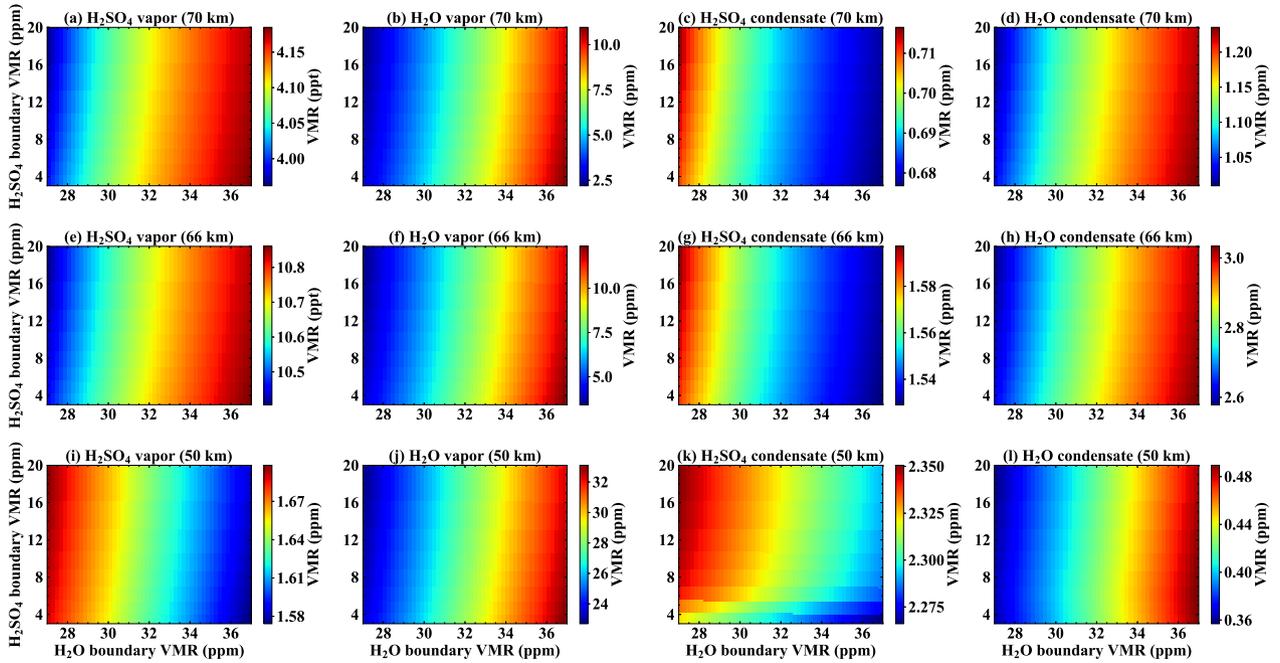}
\caption{The mixing ratios of H$_{2}$SO$_{4}$ vapor (the first column), H$_{2}$O vapor (the second column), condensed H$_{2}$SO$_{4}$ (the third column), and condensed H$_{2}$O (the fourth column) as functions of vapor mixing ratios at the lower boundary. We use 70 km (the first row), 66 km (the second row), and 50 km (the third row) to represent the cloud top, the H$_{2}$SO$_{4}$ production peak altitude, and the lower clouds, respectively.}
\label{Fig6}
\end{figure*}

\begin{figure*}
\center
\includegraphics[width=40pc]{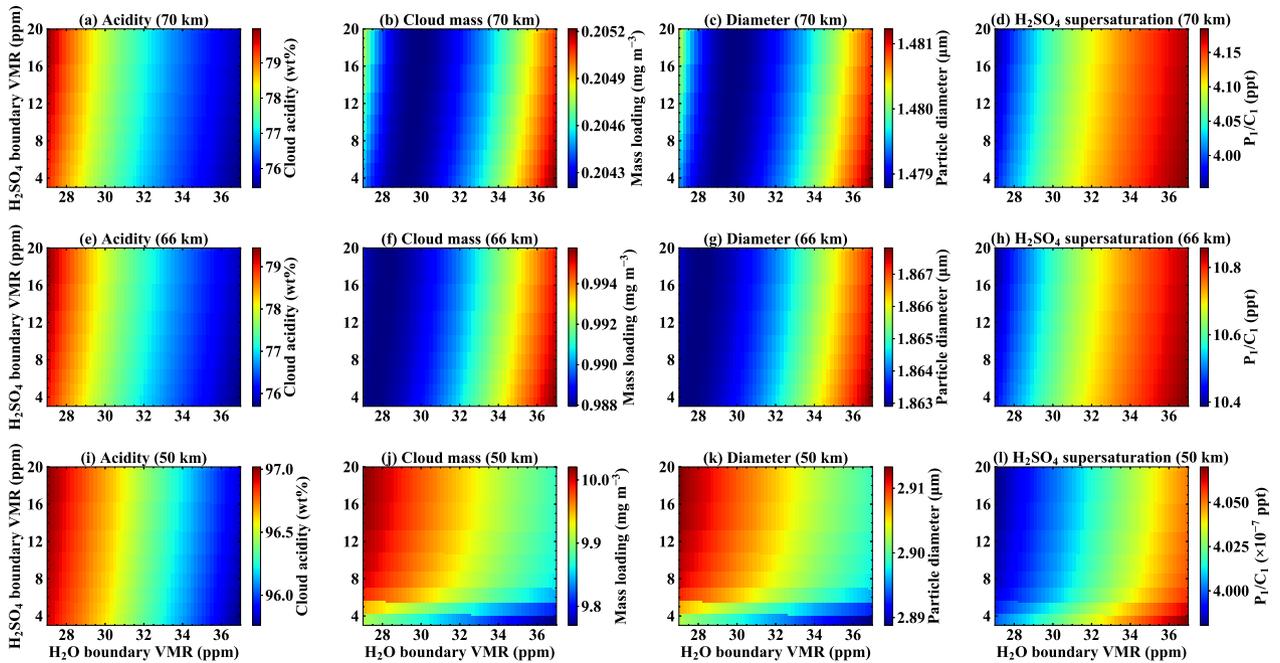}
\caption{The cloud acidity (the first column), cloud mass loading (the second column), particle diameter (the third column), and H$_{2}$SO$_{4}$ supersaturated mixing ratio (the fourth column) as functions of vapor mixing ratios at the lower boundary. The rows are the same as those in Fig.~\ref{Fig6}.}
\label{Fig7}
\end{figure*}

\subsection{Exploration of a large parameter space}
\label{sec3.2}

After the validation in Section~\ref{sec3.1}, we conclude that the semi-analytical model agrees with the D22 model pretty well. Also, the new model is much faster and allows us to explore the system behavior in a large parameter space. Here we perform the survey for different H$_{2}$SO$_{4}$ and H$_{2}$O vapor mixing ratios at the lower boundary. We vary H$_{2}$SO$_{4}$ from 3 to 20 ppm, H$_{2}$O from 27 to 37 ppm while keeping the same chemical production rate of H$_{2}$SO$_{4}$. The simulated behaviors of vapors and condensates are shown in Fig.~\ref{Fig6}, the cloud acidity, cloud mass loading, particle diameter, and H$_{2}$SO$_{4}$ supersaturation are shown in Fig.~\ref{Fig7}. We use 70 km, 66 km, and 50 km to represent the cloud top, the H$_{2}$SO$_{4}$ production peak altitude, and the lower clouds, respectively. The difference between the cloud top and the H$_{2}$SO$_{4}$ production peak altitude is small. But the responses to the boundaries are quite different between the upper clouds and the lower clouds.

In the upper cloud region, both H$_{2}$O vapor and condensed H$_{2}$O mixing ratios increase with the increase of the H$_{2}$O vapor at the lower boundary as more water vapor is diffused to the upper atmosphere. However, while the H$_{2}$SO$_{4}$ vapor mixing ratio increases, the condensed H$_{2}$SO$_{4}$ decreases as the H$_{2}$O vapor increases at the boundary. Although the variation is small, this behavior suggests an interesting response to the system. As the condensed water increases, the cloud particle size increases but the cloud acidity decreases. The former facilitates the condensation rate of H$_{2}$SO$_{4}$, but the latter decreases it (see equation~\ref{eq9}). That the condensed H$_{2}$SO$_{4}$ decreases implies that the acidity effect is more important. A smaller H$_{2}$SO$_{4}$ condensation rate leads to more H$_{2}$SO$_{4}$ vapor and less condensed H$_{2}$SO$_{4}$.

Opposite to H$_{2}$O, the upper-cloud H$_{2}$SO$_{4}$ vapor mixing ratio decreases with the increasing H$_{2}$SO$_{4}$ mixing ratio at the lower boundary, although the change is very small. The H$_{2}$SO$_{4}$ in the upper clouds is regulated by condensation and chemistry rather than diffusion. The increased H$_{2}$SO$_{4}$ at the lower boundary increases the lower-cloud condensed H$_{2}$SO$_{4}$ and cloud acidity, which increases the absorption of the H$_{2}$O vapor and weakens its upward transport. As mentioned above, a smaller H$_{2}$O flux would enhance the H$_{2}$SO$_{4}$ condensation in the upper clouds. As a result, the mixing ratio of H$_{2}$SO$_{4}$ vapor decreases, and the condensed H$_{2}$SO$_{4}$ increases.

In the lower cloud region, the H$_{2}$SO$_{4}$ vapor mixing ratio, cloud mass loading, and particle diameter show an opposite behavior to the upper clouds, while those of the H$_{2}$O vapor and condensate are similar. In the lower clouds, the H$_{2}$SO$_{4}$ vapor mixing ratio is controlled by its SVMR rather than supersaturation. As the H$_{2}$O vapor mixing ratio at the lower boundary increases, the lower-cloud acidity decreases, decreasing the H$_{2}$SO$_{4}$ SVMR. An increase of H$_{2}$SO$_{4}$ vapor mixing ratio at the lower boundary enhances the H$_{2}$SO$_{4}$ condensation and the mass loading in the lower clouds. We note that these variations are significant when the boundary H$_{2}$SO$_{4}$ vapor mixing ratio is small, as the result of the cloud base altitude increases with decreasing boundary H$_{2}$SO$_{4}$ vapor mixing ratio (see Fig.~\ref{Fig2}). 

\begin{figure*}
\center
\includegraphics[width=40pc]{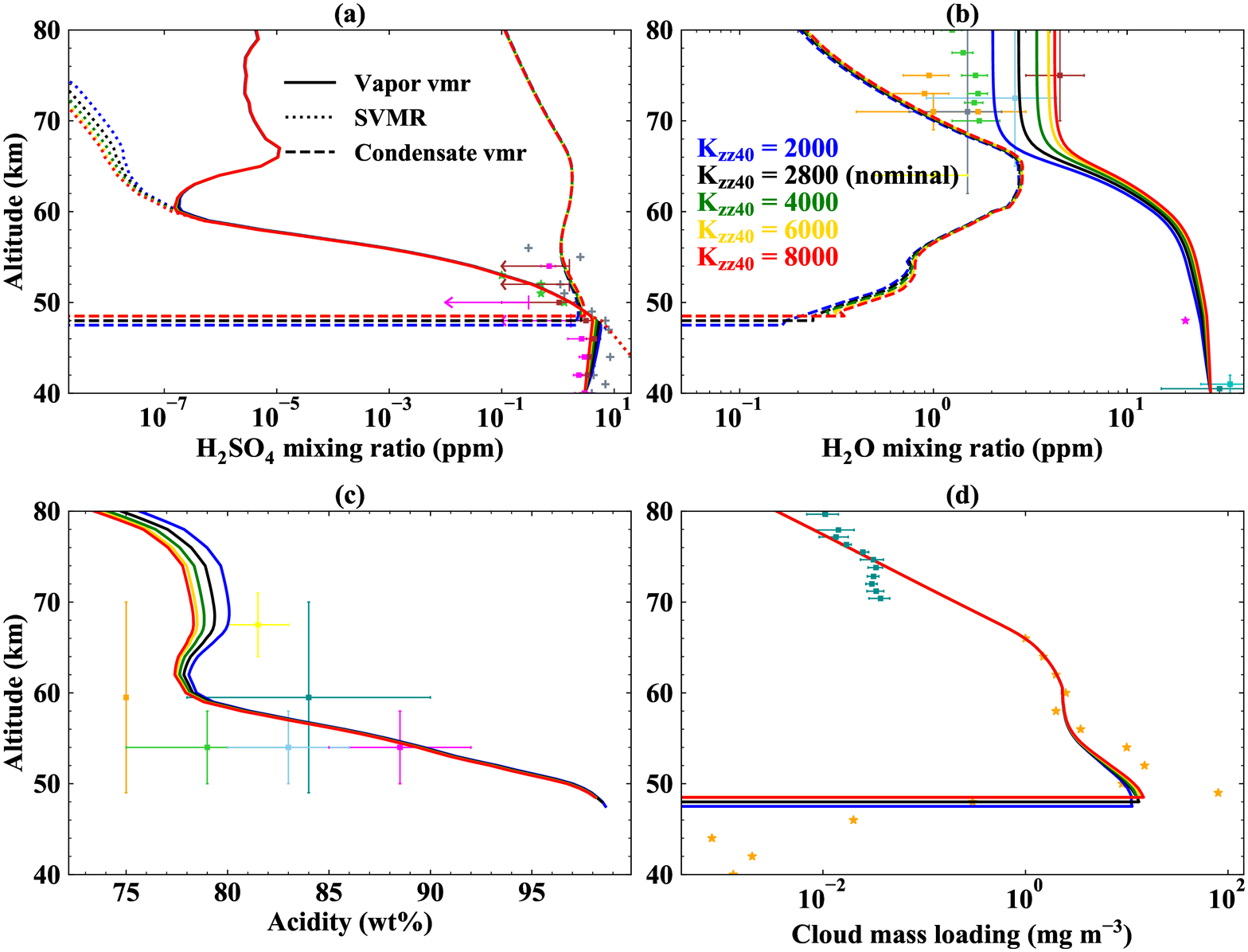}
\caption{The simulations with different eddy diffusivities in the lower atmosphere. We increase (or decrease) the standard diffusivity profile below 54 km by varying the diffusivity at 40 km to 2000, 4000, 6000, and 8000 cm$^{2}$ s$^{-1}$, respectively (see Fig.~\ref{Fig1}a). With these variations, the standard diffusivity at 48 km (near the cloud base) is increased by -13.4\%, 16.5\%, 38.6\%, and 56.8\%, respectively. We name these cases with their diffusivities at 40 km ($K_{zz40}$). The line styles, observations and estimates are the same as those in Fig.~\ref{Fig3}}
\label{Fig8}
\end{figure*}

Next, we explore the system response to varying eddy diffusivities. The diffusivity in the middle and upper atmosphere is constrained by observations \citep[][]{1983Icar...54...48L, 2021Icar..36114388M, 1981Natur.289..383W}, but it is largely unknown in the lower atmosphere. \citet[][]{1994Icar..109...58K} inferred smaller diffusivity below the clouds. Here we test different diffusivities in the lower atmosphere. We increase (or decrease) the standard diffusivity profile below 54 km by varying the diffusivity at 40 km to 2000, 4000, 6000, and 8000 cm$^{2}$ s$^{-1}$, respectively (see Fig.~\ref{Fig1}a). With these variations, the standard diffusivity at 48 km (near the cloud base) is increased by -13.4\%, 16.5\%, 38.6\%, and 56.8\%, respectively. We name these cases with their diffusivity at 40 km ($K_{zz40}$). The results are shown in Fig.~\ref{Fig8}.

With the increase of the lower atmospheric diffusivity, the cloud base is lifted. It increases the lower-cloud mass loading by strengthening the re-condensation of the H$_{2}$SO$_{4}$ vapor that evaporated from cloud-base droplets \citep[][]{2022JGRE..12707060D}. But it has a minor effect on the mass loading of the middle and upper clouds because the upper cloud mass mainly originates from the chemical production of H$_{2}$SO$_{4}$ rather than the transport from below. Moreover, with the strengthened vertical transport of H$_{2}$O vapor from the lower atmosphere, its mixing ratio increases throughout the cloud deck and the upper hazes. This is consistent with \citet[][]{2015Icar..252..327K}. Besides, the H$_{2}$O condensation is strengthened with increasing H$_{2}$O vapor mixing ratio. It also decreases the cloud acidity, especially in the upper clouds where the condensation is efficient. 

\begin{figure*}
\center
\includegraphics[width=40pc]{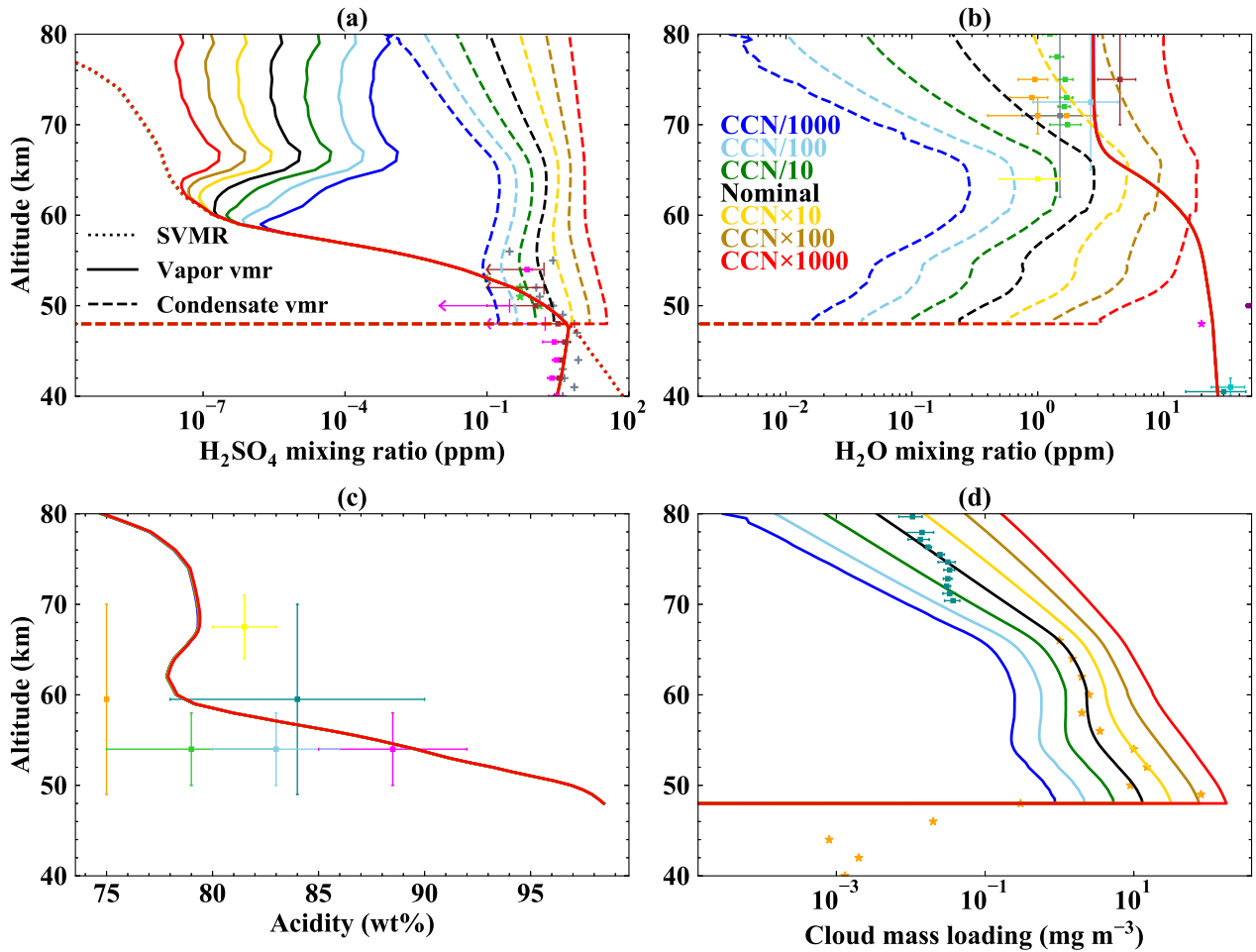}
\caption{The simulations with different CCN number density profiles. The standard CCN number density profile is scaled by -3 (blue), -2 (cyan), -1 (green), 1 (golden), 2 (light brown), and 3 (red) orders of magnitude, respectively. The nominal results (black) are represented for comparison. The line styles, observations and estimates are the same as those in Fig.~\ref{Fig3}.}
\label{Fig9}
\end{figure*}

Lastly, we explore the effect of different CCN profiles. The CCN profile is prescribed in our model since nucleation is not included, which could influence the condensation efficiency and impact the cloud formation. Our nominal CCN profile differs from observations by less than four times (see Fig.~\ref{Fig1}). But here we scale the CCN number density profile by up to three orders of magnitude to see the system response. The results are shown in Fig.~\ref{Fig9}. The increase of the CCN increases the mixing ratios of both condensates by increasing the condensation rate. Thus, the supersaturation of H$_{2}$SO$_{4}$ decreases, agreeing with \citet[][]{2022JGRE..12707060D}. Surprisingly, even if the CCN increases by three orders of magnitude, it is not able to entirely remove the supersaturation of H$_{2}$SO$_{4}$ in the upper clouds. Besides, few variations are seen in the cloud acidity and H$_{2}$O vapor because the condensation is also enhanced as the CCN number density increases.

\section{Conclusions}
\label{sec4}

This study provided a fast, semi-analytical gas-cloud model on Venus, based on the conservation of the hydrogen flux and the interaction between cloud acidity and vapor distributions for H$_{2}$SO$_{4}$ and H$_{2}$O. The physics in this model is based on the condensation cloud model in \citet[][]{2022JGRE..12707060D}. The calculation of the vapors is built on the framework of \citet[][]{1994Icar..109...58K} and \citet[][]{2015Icar..252..327K} but our model only assumes LTE for H$_{2}$O but not for H$_{2}$SO$_{4}$. Moreover, we include the feedback of cloud condensation to the vapor distributions, allowing us to explore the supersaturation of the H$_{2}$SO$_{4}$ vapor above 60 km. Similar to the KP94 model, our model can be directly solved using a semi-analytical method without time stepping. 

Our simulation agrees with the observed vertical distributions of H$_{2}$SO$_{4}$ and H$_{2}$O vapors, the cloud mass loading, and the estimate of cloud acidity. The results are consistent with \citet[][]{2022JGRE..12707060D} including a highly supersaturated H$_{2}$SO$_{4}$ layer above 60 km, different from that in the previous LTE model in \citet[][]{1994Icar..109...58K}. We confirm the conclusion in \citet[][]{2022JGRE..12707060D} that a slow condensation results in such a large vapor supersaturation in the upper cloud region. Based on the hydrogen flux conservation, we found that the vertical transport of H$_{2}$O vapor affects the cloud acidity throughout the clouds but that of H$_{2}$SO$_{4}$ vapor is only important in the lower clouds.

The fast model allows us to explore the behavior of the sulfuric acid gas-cloud system in a large parameter space. We found that the increase of H$_{2}$O vapor mixing ratio at the lower boundary decreases the H$_{2}$SO$_{4}$ condensation rate in the upper clouds, suggesting that the effect of acidity is more important than particle size on the upper-cloud H$_{2}$SO$_{4}$ condensation rate. The increase of H$_{2}$SO$_{4}$ mixing ratio at the lower boundary enhances its condensation in the lower clouds, increasing the lower-cloud mass loading. But it increases the absorption of H$_{2}$O vapor in the lower clouds and decreases its upward transport, leading to a similar behavior in the upper clouds to decreasing the H$_{2}$O at the lower boundary. Also, increasing the H$_{2}$SO$_{4}$ at the lower boundary lifts the cloud base altitude. Increasing lower atmospheric diffusivity could also lift the cloud base and increases the lower-cloud mass loading and the upward transport of H$_{2}$O vapor, but it has few effects on H$_{2}$SO$_{4}$ vapor. Increasing CCN number density by three orders of magnitude significantly decreases the H$_{2}$SO$_{4}$ supersaturation in the upper clouds but is not able to entirely remove it, further confirming that the H$_{2}$SO$_{4}$ supersaturation is robust.

On Venus, clouds provide a vital link between the upper atmosphere and the lower atmosphere. Our study shows that the observations of the cloud deck properties can be used to constrain the lower atmospheric conditions, and vice versa. Clouds are also closely connected with the atmospheric chemistry in both gas and liquids \citep[e.g.,][]{2021PSJ.....2..133R}, which need to be investigated in further detail. On extra-solar terrestrial planets, it was proposed that bright sulfuric acid clouds can be used to distinguish planets with and without the surface ocean \citep[e.g.,][]{2019ApJ...887..231L}. Since both the complicated microphysics and the time integration are avoided in our semi-analytical model, the model is 100 times faster than the previous condensation model \citep[][]{2022JGRE..12707060D} and 1000 times faster than full microphysics models \citep[e.g.,][]{2014Icar..231...83G, 2017EP&S...69..161M}. It is fast enough to be coupled with climate models and chemistry models to understand the cloudy atmospheres of Venus and Venus-like extra-solar planets.

\section*{Acknowledgements}
\label{sec5}

This work is supported by the Chinese Academy of Sciences B-type Strategic Priority Program No. XDB41000000. L. X. Zhang acknowledges support from
the National Science Foundation through grant AST1740921 and J. Cui acknowledges support from the National Natural Science Foundation of China through grant 42030201. 

\section*{Data Availability}
\label{sec6}

The data in this study is publicly available online at https://doi.org/10.6084/m9.figshare.19590562.v1 \citep[][]{Dai2022}.

\bibliographystyle{mnras}

%\bibliography{Venusian_clouds_review_1} 

\bsp	% typesetting comment
\label{lastpage}
\end{document}